\documentclass[a4paper, 10pt]{article}
\usepackage{hyperref}
\usepackage{amsmath} 
\usepackage{mathrsfs} 
\usepackage{placeins} 
\usepackage{gensymb} 
\usepackage{booktabs} 
\usepackage{graphicx} 
\usepackage[usenames, dvipsnames]{color} 
\usepackage[toc,page]{appendix} 
\usepackage[top=1in, bottom=1.25in, left=1in, right=1in]{geometry} 
\usepackage{authblk}

\usepackage{xspace}
\def\CP        {{\ensuremath{C\!P}}\xspace}
\def\SU       {{\ensuremath{\mathrm{SU}(3)}}\xspace}
\def\aSU  {{\ensuremath{\alpha_{\SU}}}\xspace}

\usepackage[normalem]{ulem}
\usepackage{color}

\def\btokpipi{B \to K \pi \pi}
\def\btokkk{B \to KK{\bar K}}
\def\babar{B{\sc a}B{\sc ar}}
\def\beq{\begin{equation}}
\def\eeq{\end{equation}}
\newcommand{\sjfit}{\mbox{\textbf{\textit{\scriptsize J}}\hspace{-0.1em}{\tiny \rm FIT}}\xspace}

 \def\PK      {\ensuremath{K}\xspace}                 
 \def\PB      {\ensuremath{B}\xspace}                 

\def\kaon    {{\ensuremath{\PK}}\xspace}
\def\Kbar    {{\kern 0.2em\overline{\kern -0.2em \PK}{}}\xspace}

\def\KorKbar    {\kern 0.18em\optbar{\kern -0.18em K}{}\xspace}

\def\Kzb     {{\ensuremath{\Kbar{}^0}}\xspace}

\def\KS      {{\ensuremath{\kaon^0_{\mathrm{ \scriptscriptstyle S}}}}\xspace}

\def\B       {{\ensuremath{\PB}}\xspace}
\def\Bbar    {{\ensuremath{\kern 0.18em\overline{\kern -0.18em \PB}{}}}\xspace}

\def\Bz      {{\ensuremath{\B^0}}\xspace}
\def\Bzb     {{\ensuremath{\Bbar{}^0}}\xspace}

\def\WorldAverageGamma {{\ensuremath{\gamma = (73.5^{+4.2}_{-5.1})\mbox{}\degree}}\xspace}

\usepackage{fancyhdr}
\title{\textbf{\boldmath Extraction of the CKM phase $\gamma$ using charmless 3-body decays of $B$ mesons}}
\author[1]{Emilie Bertholet} 
\author[1]{Eli Ben-Haim}
\author[2]{Bhubanjyoti Bhattacharya}
\author[1]{Matthew Charles}
\author[3]{David London}
\affil[1]{LPNHE, Sorbonne Universit\'e, Paris Diderot Sorbonne Paris Cit\'e, CNRS/IN2P3, Paris, France}
\affil[2]{Department of Natural Sciences, Lawrence Technological University, Southfield, MI 48075, USA}
\affil[3]{Physique des Particules, Universit\'e de Montr\'eal, \\
C.P. 6128, succ. centre-ville, Montr\'eal, QC, Canada H3C 3J7}

\begin{document}

\markright{UdeM-GPP-TH-18-265}
\maketitle
\thispagestyle{myheadings}

\begin{abstract}
The weak phase $\gamma$ is extracted from three-body charmless decays
of $B$ mesons following a method proposed by Bhattacharya, Imbeault \&
London. The result is obtained by combining the \babar\ amplitude
analyses for the processes $B^0 \rightarrow K^+ \pi^0 \pi^-$,
$B^0 \rightarrow \KS \pi^+ \pi^-$, $B^0 \rightarrow \KS \KS \KS$,
$B^0 \rightarrow K^+ \KS K^-$ and $B^+ \rightarrow K^+ \pi^+ \pi^-$,
under the assumption of \SU flavour symmetry. Six possible solutions
are found:
\begin{align*}
\begin{split}
\gamma_1	 & = [ \phantom{1} 12.9\mbox{}\,^{+8.4\phantom{1}}_{-4.3}   \text{ (stat)}  \pm 1.3 \text{ (syst)}]\degree ~,\\
\gamma_2	 & = [ \phantom{1} 36.6\mbox{}\,^{+6.6\phantom{1}}_{-6.1}  \text{ (stat)}  \pm  2.6 \text{ (syst)}]\degree ~,\\
\gamma_3	 & = [ \phantom{1} 68.9\mbox{}\,^{+8.6\phantom{1}}_{-8.6}  \text{ (stat)}  \pm  2.4 \text{ (syst)}]\degree ~,\\
\gamma_4	 & = [ 223.2\mbox{}\,^{+10.9}_{-7.5}  \text{ (stat)} \pm  1.0 \text{ (syst)}]\degree ~,\\
\gamma_5	 & = [ 266.4\mbox{}\,^{+9.2}_{-10.8}  \text{ (stat)} \pm  1.9 \text{ (syst)}]\degree ~,\\
\gamma_6	 & = [ 307.5\mbox{}\,^{+6.9\phantom{1}}_{-8.1}  \text{ (stat)}  \pm  1.1 \text{ (syst)}]\degree ~.
\end{split}
\end{align*}
One solution is compatible with the Standard Model while the others
are not. It is also found that, when
averaged over the entire Dalitz plane, the effect of \SU breaking on
the analysis is only at the percent level.

\end{abstract}

\clearpage
\section{Introduction}

In the Standard Model (SM), \CP violation in the weak sector is due to
a complex phase in the Cabibbo-Kobayashi-Maskawa (CKM) matrix.  The
CKM matrix is $3 \times 3$ and unitary; its phase information is often
represented as a triangle in the complex plane, the Unitarity
Triangle \cite{pdg}. Its three interior angles $\alpha$, $\beta$ and
$\gamma$ sum to $\pi$, and each contains information about the complex
phase. In order to test the SM, one measures $\alpha$, $\beta$ and
$\gamma$ in many different ways. Any discrepancies would suggest the
presence of physics beyond the SM.

To date, direct searches for this New Physics (NP) have not found
anything, implying that the mass scale of the NP may be beyond the
reach of present experiments. However, these new particles can still
contribute significantly to loop processes, so that flavour physics,
which is sensitive to such virtual effects, is a very promising avenue
to perform indirect searches for NP.

Of the three angles of the Unitarity Triangle, $\gamma$ is currently
the least well known: the world average value
is \WorldAverageGamma \cite{Amhis:2016xyh}. It has mainly been
extracted using processes dominated by tree-level transitions such as
$B^{\pm} \to D^{(*)} K^{(*)\pm}$ \cite{GL,GW,ADS}. One potential way
of searching for NP would therefore be to measure $\gamma$ via
loop-level processes. This can be done using charmless three-body
$B \to PPP$ decays ($P$ is a pseudoscalar meson)
\cite{diagrammatic,3body2}.

The method proposed in Ref.~\cite{3body3} uses flavour \SU symmetry
to relate $\btokpipi$ and $\btokkk$ decays. The angle $\gamma$ is then obtained
by combining information from the Dalitz plots for $\Bz \to
K^+\pi^0\pi^-$, $\Bz \to K^0\pi^+\pi^-$, $B^+ \to K^+\pi^+\pi^-$,
$\Bz \to K^+ K^0 K^-$, and $\Bz \to K^0 K^0 \Kzb$. These decay modes
all involve $b \to s$ transitions and include contributions from both
tree and penguin (loop) diagrams (the inclusion of charge-conjugate decay modes is implied throughout this paper). The extraction of $\gamma$ is
therefore potentially sensitive to NP.

A preliminary implementation of this method was carried out in
Ref.~\cite{London} using published \babar\ results. In the present paper,
we repeat the analysis of the same results, while fully taking into
account the experimental uncertainties and their correlations. We find
six possible solutions for $\gamma$. One agrees with the world-average (Standard Model) value for gamma; the other solutions do not, so that a NP scenario is allowed by the data. We are also
able to estimate the size of \SU breaking. We find that
local \SU-breaking effects can reach the usual $\sim 30\%$ level,
especially near resonances. However, when averaged over the entire
Dalitz plane, the net effect of \SU breaking is only at the percent
level.

We begin in Sec.~2 with a review of the method for extracting $\gamma$
from $\btokpipi$ and $\btokkk$ decays. Practical details of how to
implement this method are discussed in Sec.~3. In Sec.~4 we present
the results of the analysis where $\gamma$ is extracted from four
decay modes, assuming no \SU breaking. Systematic uncertainties are
discussed in Sec.~5. Sec.~6 contains a description of further tests
of \SU breaking. We conclude in Sec.~7.

\section{\boldmath Method of extraction of $\gamma$}
\label{method}

We begin with a review of the method for extracting $\gamma$ from
$\btokpipi$ and $\btokkk$ decays. There are several ingredients.
First, the amplitudes for three-body $B \to PPP$ decays can be written
in terms of diagrams \cite{diagrammatic,BPPPfullysym}. These diagrams
are similar to those of two-body decays \cite{GHLR1,GHLR2}, except
that here it is necessary to ``pop'' a quark pair from the
vacuum. Also, in contrast to two-body diagrams, the three-body
diagrams are momentum dependent.

Second, one can fix the symmetry of the final state in $B \to P_1 P_2
P_3$ by using its Dalitz plot \cite{diagrammatic}. We define the three
Mandelstam variables $s_i \equiv \left( p_j + p_k \right)^2$, where
$p_i$ is the momentum of $P_i$, and $ijk =$ 123 231 or 312. (These
obey $s_1 + s_2 + s_3 = m_B^2 + m_1^2 + m_2^2 + m_3^2$.)
Experimentally, one can reconstruct the decay amplitude
$\mathscr{A}(s_1, s_2)$, which varies as a function of position in the
Dalitz plot. The amplitude that is fully symmetric under permutations
of the final-state particles is then given by
\begin{align}
\label{symmetrsation}
\mathscr{A_{\rm{fs}}}(s_1, s_2) = \frac{1}{\sqrt{6}} \left( \mathscr{A}(s_1, s_2) + \mathscr{A}(s_2, s_1) + \mathscr{A}(s_1, s_3)
+ \mathscr{A}(s_3, s_1) + \mathscr{A}(s_3, s_2) + \mathscr{A}(s_2, s_3) \right) ~.
\end{align}
The symmetrised amplitude $\mathscr{A}_{\rm{fs}}$ has a sixfold
symmetry in the Dalitz plane. In effect, the plane can be divided
into six regions; the structure and information in each region is
identical to the others. It is therefore sufficient to consider
points in one sixth of the symmetrised Dalitz plane.

Third, in Ref.~\cite{3body2} it was shown that, as is the case in
two-body decays \cite{NR1,NR2,GPY}, under flavour \SU there are
relations between the electroweak penguin (EWP) and tree diagrams for
$b \to s$ transitions. For the fully-symmetric final state, these
take the form
\beq
\label{EWP-tree}
P'_{EWi} = \kappa T'_i ~,~~ P^{\prime C}_{EWi} = \kappa C'_i ~~ (i=1,2) ~~;~~~~
\kappa \equiv - \frac{3}{2} \frac{|\lambda_t^{(s)}|}{|\lambda_u^{(s)}|}
\frac{c_9+c_{10}}{c_1+c_2} ~,
\eeq
where the $c_i$ are Wilson coefficients and $\lambda_p^{(s)} \equiv
V^*_{pb} V_{ps}$ (the $V_{ij}$ are elements of the CKM matrix).

The method uses $\btokpipi$ and $\btokkk$ decays. In the $\btokpipi$
diagrams, the quark pair popped from the vacuum is $u{\bar u}$ or
$d{\bar d}$ (under isospin, these diagrams are equal). On the other
hand, the $\btokkk$ diagrams have a popped $s{\bar s}$ pair. Now,
under flavour \SU symmetry, which is required for the EWP-tree
relations eq.~\eqref{EWP-tree}, diagrams with a popped $s{\bar s}$
quark pair are equal to those with a popped $u{\bar u}$ or $d{\bar
d}$. In other words, under \SU the diagrams in $\btokkk$ decays are
the same as those in $\btokpipi$ decays.

Of course, flavour \SU symmetry is not exact, so one must keep track
of \SU breaking. Technically, there will be an \SU-breaking factor for
each diagram. However, if all such quantities are included, there will
be too many unknown parameters to perform a fit. For this reason, we
make the assumption that the size of \SU breaking is the same for all
diagrams, so there is a single \SU-breaking parameter \aSU relating
$\btokpipi$ and $\btokkk$ decays ($\aSU = 1$ corresponds to the
flavour-\SU limit). The idea behind this assumption is as follows. As
noted above, the diagrams are momentum dependent. This means that the
size of \SU breaking associated with a particular diagram, $\alpha_D$,
varies from point to point on the Dalitz plot. Specifically,
$\alpha_D$ will be $> 1$ at some points and $< 1$ at others. When one
averages over all points, $\alpha_D - 1$ will be small, and this will
be true for all diagrams. For this reason, we make the assumption that
the size of \SU breaking is the same for all diagrams, and we expect
it to be small. As we will see, in our fits $\aSU - 1$ is found to be
at the percent level, which supports our assumption.

Three $\btokpipi$ and two $\btokkk$ decays are used in this analysis.
They are $\Bz \to K^+\pi^0\pi^-$, $\Bz \to K^0\pi^+\pi^-$, $B^+ \to
K^+\pi^+\pi^-$, $\Bz \to K^+ K^0 K^-$, and $\Bz \to K^0 K^0 \Kzb$.
(Note that both $K^0$ and $\Kzb$ are observed as $\KS$.) As shown in
Ref.~\cite{London}, when the EWP-tree relations of
eq.~\eqref{EWP-tree} are used, the fully-symmetric amplitudes for the
five modes can be expressed as linear combinations of five effective
diagrams:
\begin{align}
\begin{split}
\label{theoreticalparams}
&2\mathscr{A_{\rm{fs}}}(B^0\to K^+ \pi^0 \pi^-) = Be^{i\gamma} - \kappa C ~,\\
& \sqrt{2}\mathscr{A_{\rm{fs}}}(B^0\to K^0 \pi^+ \pi^-) = -De^{i\gamma} - \tilde{P}'_{\rm{uc}}e^{i\gamma} - A +\kappa D ~, \\
& \mathscr{A_{\rm{fs}}}(B^0 \to K^0K^0 \overline{K}^0) = \aSU(\tilde{P}'_{\rm{uc}}e^{i\gamma} + A) ~, \\
& \sqrt{2}\mathscr{A_{\rm{fs}}} (B^0 \to K^+ K^0 K^-) = \aSU(-Ce^{i\gamma} - \tilde{P}'_{\rm{uc}}e^{i\gamma} - A +\kappa B) ~, \\
& \sqrt{2}\mathscr{A_{\rm{fs}}}(B^+\to K^+ \pi^+ \pi^-) = -Ce^{i\gamma} - \tilde{P}'_{\rm{uc}}e^{i\gamma} - A +\kappa B ~.
\end{split}
\end{align}
Here the complex parameters $A, B, C, D$ and $\tilde{P}'_{\rm{uc}}$
are linear combinations of momentum-dependent diagrams (and will, in
general, vary across the phase space), $\gamma$ is the CKM angle of
interest, and $\kappa$ is a constant defined in eq.~\eqref{EWP-tree}. As noted
above, the real quantity \aSU parametrizes the breaking of flavour \SU
symmetry and can also vary across the Dalitz plot. In the absence of
any \SU-breaking effects, $\aSU = 1$, so that the amplitudes of
$B^+\rightarrow K^+ \pi^+ \pi^-$ and $B^0 \rightarrow K^+ K^0 K^-$
modes are equal. In this limit, the fifth decay mode provides no
additional information and can be dropped from the analysis.

For each decay mode, a set of linearly-independent observables can be
formed:
\begin{align}
\begin{split}
\label{observables}
    X(s_1, s_2) & =  | \mathscr{A_{\rm{fs}}}(s_1, s_2) |^2 + | \bar{\mathscr{A}}_{\rm{fs}}(s_1, s_2) |^2 ~, \\  
    Y(s_1, s_2)  & = | \mathscr{A}_{\rm{fs}}(s_1, s_2) |^2 - | \bar{\mathscr{A}}_{\rm{fs}}(s_1, s_2) |^2 ~,\\
    Z(s_1, s_2)  & = \operatorname{Im}[ \mathscr{A}_{\rm{fs}}^*(s_1, s_2)    \bar{\mathscr{A}}_{\rm{fs}}(s_1, s_2) ] ~,
\end{split}
\end{align}
where $\bar{\mathscr{A}}_{\rm{fs}}(s_1, s_2)$ denotes the
fully-symmetric amplitude of the conjugate process. The observables
$X$, $Y$, and $Z$ are related to the effective \CP-averaged branching
fraction, the direct \CP asymmetry, and the indirect \CP asymmetry.
For a given decay, their values depend on the position in the Dalitz
plane. The observable $Z$ has no physical meaning for
flavour-specific final states such as $K^+ \pi^0 \pi^-$ and
$K^+ \pi^+ \pi^-$.

In this study, we take as experimental inputs the amplitude models
obtained by \babar\ in Refs.~\cite{KPP,KsPP,KPP0,KsKsKs,KsKK}.
The \babar\ analysis of $\Bz \rightarrow \KS \KS \KS$~\cite{KsKsKs} was time-integrated
and \CP-averaged; since no distinction was made between $B_0$ and
$\overline{B}_0$, only the observable~$X$ is accessible for this mode.
As noted in Ref.~\cite{London}, this implies a simplification in the
expression for its amplitude compared with
eq.~\eqref{theoreticalparams}. To be specific, the requirement that $Y
= Z = 0$ implies that $\tilde{P}'_{\rm{uc}} = 0$, so that
\begin{equation}
  \label{eq:theoreticalparamsForKSKSKS}
  \mathscr{A_{\rm{fs}}}(B^0 \rightarrow K^0K^0 \bar{K}^0) = \aSU A ~.
\end{equation}

Since for each mode the observables $X,Y,Z$ depend upon the
fully-symmetric amplitude $\mathscr{A_{\rm{fs}}}$ [Eq.~\eqref{observables}], and $\mathscr{A_{\rm{fs}}}$ is related to
the theory parameters by Eqs.~\eqref{theoreticalparams}
and~\eqref{eq:theoreticalparamsForKSKSKS}, the observables may be
written as functions of those theoretical parameters. Expressing them
in terms of magnitudes and strong phases ($U=u e^{i\phi_{u}}$ for
$U=A,B,C,D$), and setting $\phi_{a} = 0$ without loss of generality,
the following relations are obtained:
\begin{align}
\begin{split}
\label{XYZ}
    &X^{th}_{K^+ \pi^+ \pi^-}(s_1, s_2) = a^2+(\kappa b)^2 + c^2 + 2 ac \cos\phi_c \cos\gamma - 2 \kappa a b \cos\phi_b - 2 \kappa b c \cos(\phi_b-\phi_c)\cos\gamma ~, \\  
    &Y^{th}_{K^+  \pi^+ \pi^-}(s_1, s_2) = -2 \left( a c \sin\phi_c+\kappa b c \sin(\phi_b-\phi_c) \right)\sin\gamma ~, \\
    &X^{th}_{\KS K^+ K^-}(s_1, s_2) = \aSU^2 X^{th}_{K^+ \pi^+ \pi^-} ~, \\  
    &Y^{th}_{\KS K^+ K^-}(s_1, s_2) = \aSU^2 Y^{th}_{K^+  \pi^+ \pi^-} ~, \\
    &Z^{th}_{\KS K^+ K^-}(s_1, s_2) = \aSU^2 \left(-c^2 \cos\gamma - a c \cos\phi_c + \kappa b c \cos(\phi_b-\phi_c) \right)\sin\gamma ~, \\
    &X^{th}_{\KS \pi^+ \pi^-}(s_1, s_2) = a^2+(\kappa d)^2 + d^2 + 2 ad \cos\phi_d \cos\gamma - 2 \kappa a d \cos\phi_d - 2 \kappa d^2 \cos\gamma ~, \\  
    &Y^{th}_{\KS \pi^+ \pi^-}(s_1, s_2) = -2 a d \sin\phi_d \sin\gamma ~, \\
    &Z^{th}_{\KS \pi^+ \pi^-}(s_1, s_2) = \left(-d^2 \cos\gamma - a d \cos\phi_d + \kappa d^2 \right) \sin\gamma ~, \\
    &X^{th}_{K^+ \pi^+ \pi^0}(s_1, s_2) = \frac{1}{2} \left( b^2 + \kappa^2 c^2 - 2\kappa b c \cos \gamma \cos(\phi_{b}-\phi_{c}) \right) ~, \\  
    &Y^{th}_{K^+  \pi^+ \pi^0}(s_1, s_2) = \kappa b c \sin \gamma \sin(\phi_b - \phi_c) ~, \\
    &X^{th}_{\KS \KS \KS}(s_1, s_2) = 2 \aSU^2 a^2 ~.
\end{split}
\end{align}

If $\gamma$ is extracted at a single point $(s_1, s_2)$ on the Dalitz
plane, there are nine real, unknown parameters: four magnitudes ($a$,
$b$, $c$, $d$), three strong phases ($\phi_{b}$, $\phi_{c}$,
$\phi_{d}$), $\gamma$, and \aSU. From the experimental input, there
are eleven observables: three ($X,Y,Z$) for each of the modes $\KS K^+
K^-$ and $\KS \pi^+ \pi^-$, two ($X,Y$) for each of the modes
$K^+ \pi^+ \pi^-$ and $K^+ \pi^+ \pi^0$, and one ($X$) for $\KS \KS
\KS$. If \aSU is fixed to unity, there are instead eight unknown
parameters and nine observables. In both cases, there are more
observables than theory parameters, and $\gamma$ may be extracted with
a fit.

One can instead determine $\gamma$ using information from several
points on the Dalitz plane simultaneously. This increases the number
of observables, but it also increases the number of unknowns, since
all those parameters that describe the strong dynamics of the decay
can vary across the Dalitz plane (whereas $\gamma$ of course does
not). Thus, if $N$ points on the Dalitz plane are used, there are
$11N$ observables and $8N+1$ unknown parameters when \aSU is allowed
to vary, or $9N$ observables and $7N+1$ unknowns if \aSU is fixed to
unity. For any $N \geq 1$, the number of observables exceeds the
number of unknowns in both cases, so that $\gamma$ may again be
extracted with a fit.

\section{Practical implementation of the extraction method}
\label{implementation}

The five \babar\ analyses use the isobar formalism to parametrise the
variation of the amplitude across the Dalitz plane. In this approach,
the total amplitude at a point $(s_{1}, s_{2})$ in the Dalitz plane is
given by the coherent sum of the amplitudes of $n$ individual
decay channels:
\begin{align}
\label{amp}
A(s_{1}, s_{2}) = \sum_{j=1}^{n} c_j  F_j(s_{1}, s_{2}) ~,
\end{align}
where the isobar coefficients, $c_j$, are complex numbers containing all the weak phase dependence, and the lineshapes, $F_j$, are wave functions
(such as Breit--Wigner functions) that describe the dynamics of the decay amplitudes.

The isobar coefficients and the lineshapes given in \babar's papers
are used to compute the amplitudes of the different decay modes as a
function of position in the Dalitz plane. This calculation is
implemented with the LAURA$^{++}$ software package~\cite{Laura++}.
The uncertainties on the isobar coefficients quoted by \babar, and the
associated correlation matrices, are used to compute the experimental
uncertainties on the extracted values of the angle $\gamma$. For the
decay mode $B^0 \to \KS \KS \KS$, no correlation matrix was quoted,
and the correlations are therefore neglected.

After encoding the isobar model for a decay mode in LAURA$^{++}$, the
amplitudes for the process and its \CP conjugate may be calculated
at any point, or set of points, on the Dalitz plot. From the
amplitudes, those observables ($X$, $Y$, $Z$) that are well-defined
may be computed according to eq.~\eqref{observables}. The uncertainties
on the isobar coefficients are propagated to obtain the uncertainties
on the observables along with their correlations.

It is also possible to compute the observables from the theory
parameters eq.~\eqref{XYZ}; for a set of $N$ points, those theory
parameters consist of $\gamma$ plus $N$ instances of the amplitude and
\SU-breaking parameters, $\{ a, b, c, d, \phi_b, \phi_c, \phi_d
\rm{\,and\,} \aSU \}$.  The $\chi^2$ may then be computed for
compatibility between the observables expected given these theory
parameters, and the observables obtained from experimental inputs. To
constrain $\gamma$, a $\chi^2$ scan is performed as follows: $\gamma$
is fixed to a certain value, then a $\chi^2$ minimisation is performed
with the other $8N$ parameters free to vary.
In principle, for a global minimisation the final values of the fit
parameters and the $\chi^2$ should not depend on the initial values of
the parameters. In practice, for multidimensional fits some
dependency is observed due to the presence of secondary local minima.
In order to obtain a more robust estimate of the global minimum the
minimisation process is repeated 500 times, with values of the initial parameters varied randomly in a large physical range each time. Then,
for each fixed value of $\gamma$, the smallest $\chi^2$ is retained.
The value of $\gamma$ is then increased by one step and the
minimisation repeated. Performing this many times, a scan of the
$\chi^2$ as a function of $\gamma$ is obtained.
The minima of this scan are the preferred values for $\gamma$. The procedure for finding the minima in a scan is detailed in Appendix~\ref{appendixA}. The
asymmetric statistical uncertainty on each solution is then estimated
as the change in $\gamma$ required to produce a change of one unit in
$\chi^2$ from the minimum.

Flavour \SU symmetry can be broken locally when considering single
points on the Dalitz plane. However, as shown in Sec.~\ref{SU3}
below, \SU-breaking effects are small when averaging over a large
number of points. For this reason, as well as to minimise the
statistical uncertainties and to use the maximum amount of information
possible, it is desirable to extract $\gamma$ using the largest
possible number of points. In principle, $\gamma$ could be obtained
from an arbitrarily large set of points. However, the observables can
be highly correlated between points, especially if the same resonance
(or non-resonant component) is the dominant contributor to the points
in one of the decay modes. High correlations have an impact on the
covariance matrix, which becomes approximately singular and not
invertible. This imposes practical limitations on the choice of
points: the total number of points that can be used in a fit is finite
and small. For these modes and isobar models, it is found that no more
than three points can be used, and not all three-point combinations
are possible. In order to avoid dependence on the choice of points
(i.e., to avoid experimenter's bias), the scan procedure is carried out
repeatedly with random combinations of three points, applying a filter
to reject sets of points for which the correlation matrix contains
entries above 70\%. In total, 501 random sets of three points that
pass this filter are used.

For each of the solutions (scan minima) for $\gamma$, the final result
is taken to be the average over the central values for that solution,
and the uncertainty on that result is the average of the uncertainties
from individual scans. Note that, fluctuations aside, the average
uncertainty does not decrease as more scans are added.

As discussed in Sec.~\ref{method}, the extraction of $\gamma$ can
be performed with four modes (fixing $\aSU=1$ and so neglecting
flavour \SU symmetry-breaking effects), or with five modes (allowing
\aSU to vary and parametrising those effects). For reasons of fit
stability and convergence, the method with four modes is chosen to be
the baseline for the results and will be presented in the next
section. The method using five modes is then used to assess the
systematic uncertainties associated with \SU breaking; the
corresponding uncertainties are given in Sec.~\ref{systematics}, with
the results for the five-mode method described in more detail in
Appendix~\ref{5modessym1}.

\section{\boldmath Results with four modes, $\alpha_{SU(3)} = 1$}
\label{results}
\label{4modessym1}

The method described in the previous section is applied, and $\chi^2$ scans for $\gamma$ are obtained. Six distinct minima are found; 
averaging over the 501 sets of three points in the Dalitz plot,
their central values $\mu$ and asymmetric experimental uncertainties ($\sigma_L$, $\sigma_R$) are given in Table~\ref{Minima_sym1_4modes}. The experimental uncertainties are below $11\degree$ in each case. The third of these minima, at $68.9\degree$, is compatible with the current world-average value of $\gamma$ \cite{Amhis:2016xyh}.

\begin{figure}
\begin{center}
    \includegraphics[width=0.6\textwidth]{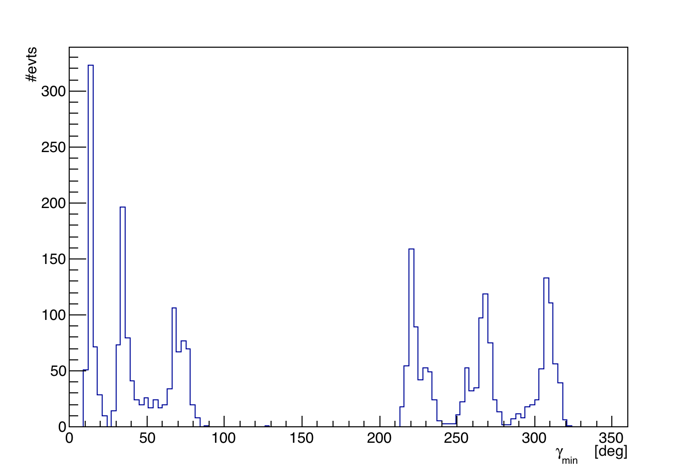}
\end{center}
\caption{The minima found with four decay modes ($\aSU=1$). For each of the 501 sets of random combinations of three points in the Dalitz plot, a $\chi^2$ scan for $\gamma$ is performed and the minima $\gamma_{\mathrm{min}}$ are found. The histogram shows the accumulation of the minima across all 501 scans.}
	\label{minima4modesSym1_fit}
\end{figure}

Since each combination of three points carries different information,
the form of the $\chi^2$ scans vary from one combination of points to the next.
The central values fluctuate, and, in some instances, not all of the six minima are present.
The distribution of the minima across the scans is shown in
Figure~\ref{minima4modesSym1_fit},
and the rates at which the minima are found are given
in Table~\ref{freq_4modes};
each of them is found in more than 90\% of scans.

\begin{table}
\caption{
  The minima found with four decay modes ($\aSU=1$).
  For each minimum, the central value for $\gamma$ is given ($\mu$),
  along with the asymmetric experimental uncertainty on the left-
  and right-hand sides ($\sigma_L$, $\sigma_R$).
  }
\label{Minima_sym1_4modes}
\begin{center}
\begin{tabular}{crrr}
 &
\multicolumn{1}{c}{$\mu$} &
\multicolumn{1}{c}{$\sigma_L$} &
\multicolumn{1}{c}{$\sigma_R$}  \\
\midrule
Minimum 1 & $ 12.9\degree$ & $ 4.3\degree$ & $ 8.4\degree$ \\
Minimum 2 & $ 36.6\degree$ & $ 6.1\degree$ & $ 6.6\degree$ \\
Minimum 3 & $ 68.9\degree$ & $ 8.6\degree$ & $ 8.6\degree$ \\
Minimum 4 & $223.2\degree$ & $ 7.5\degree$ & $10.9\degree$ \\
Minimum 5 & $266.4\degree$ & $10.8\degree$ & $ 9.2\degree$ \\
Minimum 6 & $307.5\degree$ & $ 8.1\degree$ & $ 6.9\degree$ \\
\end{tabular}
\end{center}
\end{table}

\begin{table}
\caption{The rates at which the different minima are obtained with four decay modes ($\aSU=1$). A total of 501 scans are used.}
\label{freq_4modes}
\begin{center}
\begin{tabular}{lcc}
 & Count  & Fraction (\%)  \\
\midrule
Minimum 1	 & 	484	     & 	 96.6   \\
Minimum 2	 & 	474	     & 	 94.6	\\
Minimum 3	 & 	461	     & 	 92.0    \\
Minimum 4	 & 	499	     & 	 99.6	 \\
Minimum 5	 & 	487	     & 	 97.2	 \\
Minimum 6	 & 	488	     & 	 97.4	  \\
\end{tabular}
\end{center}
\end{table}

\FloatBarrier
\section{Systematic uncertainties}
\label{systematics}

The experimental statistical and systematic uncertainties
on the amplitude models used as inputs
are already included in the results
given in Table~\ref{Minima_sym1_4modes}.
Two additional sources of systematic uncertainty,
discussed below,
are considered in this study.
The first relates to the combination of the minima
obtained with different sets of three points in the Dalitz plot.
The second relates to flavour \SU breaking.
The results are summarised in table \ref{Systematics}. 

The form of the $\chi^2$ scan varies according to the points chosen,
and in some instances a minimum is found successfully
but is not well separated from another nearby minimum,
such that if the two minima are at $\mu_1$ and $\mu_2$
with $\chi^2$ values $\chi^2_1$ and $\chi^2_2$,
no value of $\gamma$ in the range $\mu_1 < \gamma < \mu_2$
(or $\mu_2 < \gamma < \mu_1$)
has a $\chi^2$ value greater than or equal to $\chi^2_1+1$.
This means that the algorithm
set out in Sec.~\ref{implementation}
cannot determine the
experimental uncertainty on $\mu_1$.
These minima are referred to as poorly resolved
and are not included in the average from which the
overall results are obtained
(Table~\ref{Minima_sym1_4modes}).
Discarding these minima could have a systematic
effect on the average
(e.g., if $\mu_1 < \mu_2$ are two nearby minima
then upward fluctuations in $\mu_1$
are more likely to be too close to $\mu_2$ to resolve
than downward fluctuations in $\mu_1$,
potentially causing a negative bias in $\mu_1$).
To assess this effect, the analysis is repeated
including all minima from all scans in the average,
even those that are not well resolved.
The systematic uncertainty is then assessed as
\begin{align}
\label{syst1}
\sigma_{\mathrm{poorly~resolved}} = |\mu-\mu^{\mathrm{all}}| ~,
\end{align}
where
$\mu$ is the central value obtained
including only well-resolved minima in the average, and
$\mu^{\mathrm{all}}$ is the central value obtained when including both
well-resolved and not-well-resolved minima.
The values obtained are given in
Table~\ref{Systematics} and are below $1.5\degree$ for each minimum.

The extraction performed with four modes does not take into
account flavour \SU breaking. While it is not practical
to allow for \SU breaking in a completely general way
in this analysis, the scale of the effect can be assessed
by allowing the \SU-breaking parameter \aSU
to vary and seeing how much the values of $\gamma$ change.
To this end, the analysis is repeated using five modes instead
of four, and with \aSU free to vary as an
additional real parameter in the fit. As before, a $\chi^2$
scan for $\gamma$ is obtained with hundreds of random combinations
of three points in the Dalitz plot, and for each scan the
minima are found. (More details are given in Appendix~\ref{5modessym1}.)
For each minimum, the central value of $\gamma$ is averaged
over the scans as before. These estimates using five modes
($\mu^{5\,\mathrm{modes}}$)
may then be compared to the value for that minimum
obtained with the baseline, four-mode procedure ($\mu$)
to assess how large an effect flavour \SU breaking
has on the value of $\gamma$:
\begin{align}
\label{syst2}
\sigma_{\SU} = |\mu-\mu^{5\,\mathrm{modes}}| ~.
\end{align}
The values obtained are given in
Table~\ref{Systematics} and are below $3\degree$ for each minimum.
More tests of the validity of the flavour \SU symmetry
hypothesis are described in Sec.~\ref{SU3}.

\begin{table}[h!]
\caption{Summary of the systematic uncertainties.}
\label{Systematics}
\begin{center}
\begin{tabular}{lcc}
 & Poorly resolved minima  & Flavour \SU breaking  \\
\midrule
Minimum 1	 & 	$0.8\degree$     & 	 $1.0\degree$   \\
Minimum 2	 & 	$0.3\degree$     & 	 $2.6\degree$	\\
Minimum 3	 & 	$0.2\degree$     & 	 $2.4\degree$    \\
Minimum 4	 & 	$0.7\degree$     & 	 $0.7\degree$	 \\
Minimum 5	 & 	$1.4\degree$     & 	 $1.3\degree$	 \\
Minimum 6	 & 	$0.7\degree$     & 	 $0.9\degree$	  \\
\end{tabular}
\end{center}
\end{table}

\FloatBarrier
\section{\boldmath Studies of flavour \SU breaking}
\label{SU3}

The assumption of flavour \SU, and specifically that 
$\aSU = 1$, is tested in two further ways.
The first involves comparing the amplitudes of two modes
related by flavour \SU as a function of position
in the Dalitz plane.
The second consists of determining the value of
\aSU over the Dalitz plane
from fits to the amplitude models.

\subsection{\boldmath Comparison of the amplitudes of $B^0 \to K_S K^+ K^-$ and $B^+ \to K^+ \pi^+ \pi^-$}
\label{alpha1}

From inspection of the last two lines of
eq.~\ref{theoreticalparams}, there is a linear relationship
between the fully symmetric amplitudes for
$B^0 \to  \KS K^+ K^-$  and $B^+ \to  K^+ \pi^+ \pi^-$:
\begin{equation}
\mathscr{A}_{\rm{fs}} (B^0 \rightarrow K^+ K^0 K^-) = \aSU \, 
\mathscr{A}_{\rm{fs}} (B^+ \rightarrow K^+ \pi^+ \pi^-) ~.
\end{equation}
The value of the parameter \aSU, a measure of the amount of the local
flavour \SU breaking, can be inferred by comparing the values of the
amplitudes of these modes at different points on the Dalitz
plane \cite{BPPPfullysym}. We define the following ratio:
\begin{align}
R(s_{13}, s_{23})
=
\left| \frac{
\mathscr{A}_{\rm{fs}} (B^+ \rightarrow K^+ \pi^+ \pi^-; s_{13}, s_{23})
+
\mathscr{A}_{\rm{fs}} (B^- \rightarrow K^- \pi^- \pi^+; s_{13}, s_{23})
}{
\mathscr{A}_{\rm{fs}} (\Bz \rightarrow K^+ \KS   K^-  ; s_{13}, s_{23})
+
\mathscr{A}_{\rm{fs}} (\Bzb \rightarrow K^- \KS   K^+  ; s_{13}, s_{23})
}
\right|
\end{align}
where $\mathscr{A}_{\rm{fs}}(X;s_{13}, s_{23})$
is the symmetrised amplitude for the decay mode $X$
measured at point $(s_{13}, s_{23})$.
The ratio is an estimate of \aSU at that point.

\begin{figure}[bthp]
  \begin{center}
  \begin{tabular}{cc}
    \includegraphics[width=.48\textwidth]{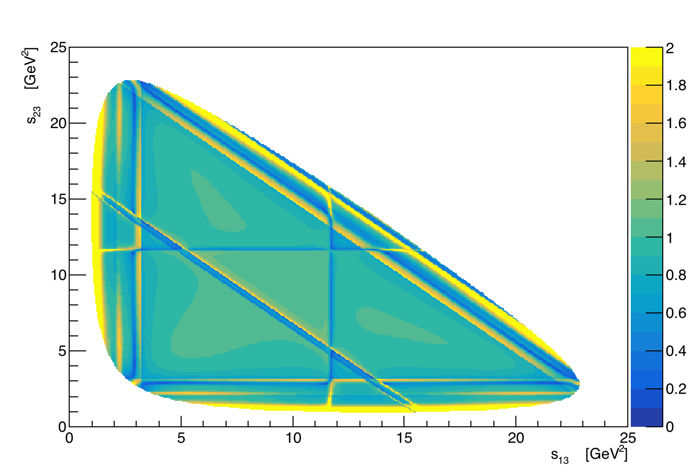} &
    \includegraphics[width=.48\textwidth]{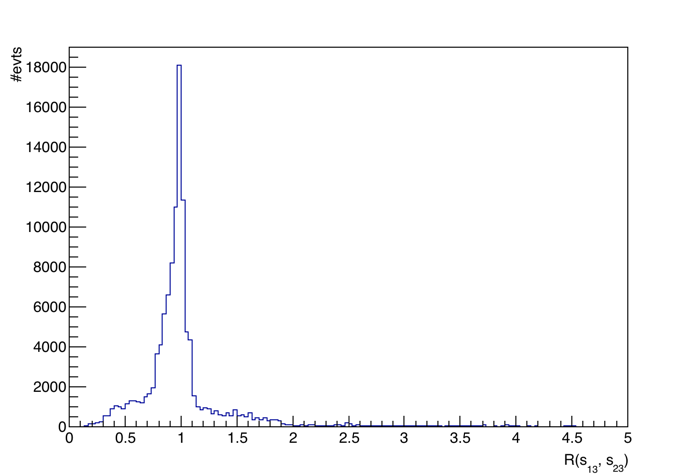} \\
    \small (a) &
    \small (b)
  \end{tabular}
  \end{center}
  \caption{(a) Ratio of amplitudes $R(s_{13}, s_{23})$ over the whole fully symmetrised DP. Note that the $R(s_{13}, s_{23})$ scale is truncated at 2.0. (b) Histogram of the different values of the ratio of amplitudes $R(s_{13}, s_{23})$.}
\label{alpha_mapping}
\end{figure}

Figure~\ref{alpha_mapping}\,(a) shows the value of $R(s_{13}, s_{23})$
as a function of position in the Dalitz plane.
Significant deviations from unity are seen, especially near resonances.
This is unsurprising, given that flavour \SU is broken by the
mass difference between $s$ and $u/d$ quarks.
A histogram of the values of $R$, sampled uniformly across the Dalitz plane,
is shown in Fig.~\ref{alpha_mapping}\,(b).
The distribution peaks near one, and the average value is $1.028$,
rather close to unity. This suggests qualitatively that, while
\SU is strongly violated locally, it holds reasonably well
when averaging across the phase space.

\subsection{\boldmath Fitted value of \aSU over the Dalitz Plane}

Another approach is to determine \aSU from a fit.
For this exercise, individual points in the Dalitz plane are
considered (as opposed to sets of three points).
A uniform grid of 386 points is used.
For each point, a similar procedure is followed to that described
in Sec.~\ref{implementation}, with a $\chi^2$ minimisation carried out
with $\gamma$ being fixed to a certain value $\gamma_i$ and the other physics
parameters, including \aSU, being free to vary.
As before, the fit is repeated 500 times (for each point)
with the initial parameter
values randomised, and the solution with the smallest $\chi^2$
after the fit is retained.
However, instead of scanning for $\gamma$ across the full range,
the exercise is only performed for six values $\gamma_i$
corresponding approximately to the six minima given in
Table~\ref{Minima_sym1_4modes}.
At each point and for each value of $\gamma_i$ tested,
the value of \aSU for the best-fit solution is recorded.

\begin{figure}[bthp]
 	\begin{center}
    \includegraphics[width=0.6\textwidth]{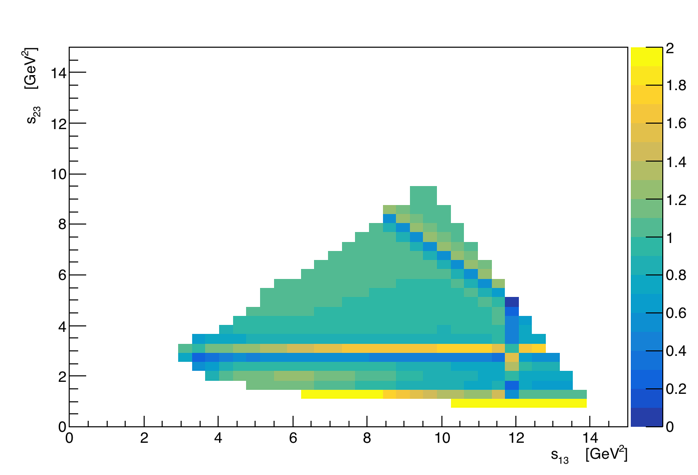}
    \end{center}
 	\caption{Histogram of the extracted values of \aSU
    at each point of a grid of 386 points uniformly spaced
    and covering the symmetrised Dalitz plane.
    The results are averaged over the six $\gamma_i$. Note that the $R(s_{13}, s_{23})$ scale is truncated at 2.0.
	}
	\label{AlphaSU3avg}
\end{figure}

Averaging over the uniform grid of points,
the mean values of \aSU are given in
Table~\ref{talbeAlpha} for each $\gamma_i$.
Each value is close to unity, and
negligible variation in the average \aSU
is seen between the six minima.
The variation of \aSU with position in the Dalitz plot
is illustrated in Fig.~\ref{AlphaSU3avg},
in which at each point in the symmetrised Dalitz plot the
fitted values of \aSU from the six $\gamma_i$
are averaged. Similar structure is seen to that
observed in
Fig.~\ref{alpha_mapping}, and
the breaking of flavour \SU is clearly seen near resonances.

  \begin{table}[bthp]
  \caption{
    The six values $\gamma_i$ assumed when investigating
    the variation of \aSU across the Dalitz plane, and the
    average value of \aSU obtained for each,
    $\langle \aSU \rangle_i$.
  }
\begin{center}
\begin{tabular}{c|c}
  $\gamma_i$ & $\langle \aSU \rangle_i$  \\
  \midrule
  12\degree   & 1.06 	 \\
  37\degree	 & 1.06	 \\
  68\degree	 & 1.05	 \\
  223\degree  & 1.06 \\
  266\degree  & 1.05	 \\
  307\degree  & 1.05	  \\
  \end{tabular}
\end{center}
\label{talbeAlpha}
  \end{table}

\section{Conclusion}

The method of extracting the weak phase $\gamma$ from three-body
charmless decays of the $B$ meson developed by Bhattacharya, Imbeault
and London~\cite{London} is applied to amplitude models of five
charmless three-body decays of $B$ mesons obtained by the \babar\
collaboration~\cite{KPP,KsPP,KPP0,KsKsKs,KsKK}. Six solutions for
$\gamma$ are found:
\begin{align*}
\begin{split}
\gamma_1	 & = [ \phantom{1} 12.9\mbox{}\,^{+8.4\phantom{1}}_{-4.3}   \text{ (stat)}  \pm 1.3 \text{ (syst)}]\degree,\\
\gamma_2	 & = [ \phantom{1} 36.6\mbox{}\,^{+6.6\phantom{1}}_{-6.1}  \text{ (stat)}  \pm  2.6 \text{ (syst)}]\degree,\\
\gamma_3	 & = [ \phantom{1} 68.9\mbox{}\,^{+8.6\phantom{1}}_{-8.6}  \text{ (stat)}  \pm  2.4 \text{ (syst)}]\degree,\\
\gamma_4	 & = [ 223.2\mbox{}\,^{+10.9}_{-7.5}  \text{ (stat)} \pm  1.0 \text{ (syst)}]\degree,\\
\gamma_5	 & = [ 266.4\mbox{}\,^{+9.2}_{-10.8}  \text{ (stat)} \pm  1.9 \text{ (syst)}]\degree,\\
\gamma_6	 & = [ 307.5\mbox{}\,^{+6.9\phantom{1}}_{-8.1}  \text{ (stat)}  \pm  1.1 \text{ (syst)}]\degree ~.
\end{split}
\end{align*} 
The six values obtained are well separated, and one is compatible with
the Standard Model while the others are not. The central values and
statistical uncertainties are obtained under the hypothesis of \SU
symmetry; the systematic uncertainties indicate the effect of flavour
\SU breaking as well as the impact of poorly resolved minima on the
procedure.  The statistical uncertainty is dominant, and is below
$11\degree$ for each of the six solutions. This is approximately a
factor two larger than the uncertainty on the world-average value of
$\gamma$, and allows the value obtained from these loop-level
processes to be compared to the tree-dominated average. The presence
of multiple solutions may reflect trigonometric ambiguities in the
amplitudes.

Further tests of the flavour \SU symmetry hypothesis were performed,
studying the variation in the \SU-breaking parameter \aSU across the
phase space. Strong local variation is seen, comparable to the $\sim
30\%$ level typically considered, but the average value of \aSU is
found to be close to 1 (corresponding to \SU symmetry) within a few
percent.

The study presented in this paper is a complete proof of principle,
including fully-propagated experimental uncertainties. It would
benefit from additional and more precise experimental inputs; results
from Belle~II and LHCb would be welcome. It is worth noting that
certain modes are well suited to the LHCb detector (e.g. $B^+ \to K^+
\pi^+ \pi^-$), while others are better adapted to Belle~II (e.g. $B^0
\to \KS \KS \KS$). Given this, one interesting possibility would be a
simultaneous fit of the physics parameters to datasets of both
experiments using a framework such as \sjfit~\cite{Ben-Haim:2014afa}.

Further developments on the theoretical side would also be welcome,
such as considering other symmetry states (fully antisymmetric or of
mixed symmetry). This would add information, thereby reducing the
statistical uncertainties, and might help to resolve the ambiguities
and determine whether the value of $\gamma$ found using loop-level
processes is or is not equal to that obtained using tree-level decays.

\section*{Acknowledgements}

We would like to thank Maxime Imbeault for discussions and collaboration during the early stages of this project. The work of B. B. was supported by Lawrence Technological University through a faculty seed grant. The work of D. L. was financially supported in part by NSERC of Canada.

\clearpage
\begin{appendices}
\section{Algorithm for extracting the minima}
\label{appendixA}

The following algorithm is used to find the minima in a given scan:

\begin{enumerate}
\item Start at the first point.
\item Define the current window to be the range of $\gamma$ spanned by the
  current point plus the next 19 consecutive points.
  Fit those 20 points with a 3$^{\mathrm{rd}}$-order polynomial function.
\item Determine the minimum of the fitted polynomial
  (at $x =\gamma_{\rm{min}}$, $y = \chi^2(\gamma_{\rm{min}})$
\item Reject the minimum ($\gamma_{\rm{min}}, \chi^2(\gamma_{\rm{min}})$) if any of the following is true:
    \begin{itemize}
    \item The value of $\gamma_{\rm{min}}$ is outside the window.
    \item $\chi^2(\gamma_{\rm{min}}) > 7$.
    \item The polynomial fit is of poor quality (its fit $\chi^2$ is greater than 5).
    \end{itemize}
\item Move along one point, then go back to step 2 (unless the points have been exhausted).
\end{enumerate}

Usually, when a minimum is identified,
it will be found by several consecutive polynomial fits (steps 2--4).
Due to statistical fluctuations, the value of $\gamma_{\rm{min}}$
will differ slightly between these; the average value is taken.

\FloatBarrier
\section{\boldmath Extraction of $\gamma$ with five modes varying \aSU in the fit}
\label{5modessym1}

The analysis (described in Sec.~\ref{implementation})
was carried out using four decay modes and with \aSU
fixed to unity as a baseline.
To assess the systematic effect of \SU breaking, a similar
procedure was used with five decay modes and with \aSU
free to vary in the fit. The following changes were made to the procedure: the rejection criterion on the correlation between sets of points was relaxed from 70\% to 80\%, and the number of random set of three points was reduced from 501 to 401.
The fit behaviour was found to be less stable, with convergence of the
$\chi^2$ minimisation in around 80\% of cases (rather than 100\% in the
baseline). The frequency with which the minima were identified was also
reduced (as shown in Table~\ref{freq_5modes}). The reduced stability
is taken to be due to the increased number of free parameters, and the
consequent increase in the size of the covariance matrix.

\begin{figure}[bthp]
 	\centering
    \includegraphics[width=0.6\textwidth]{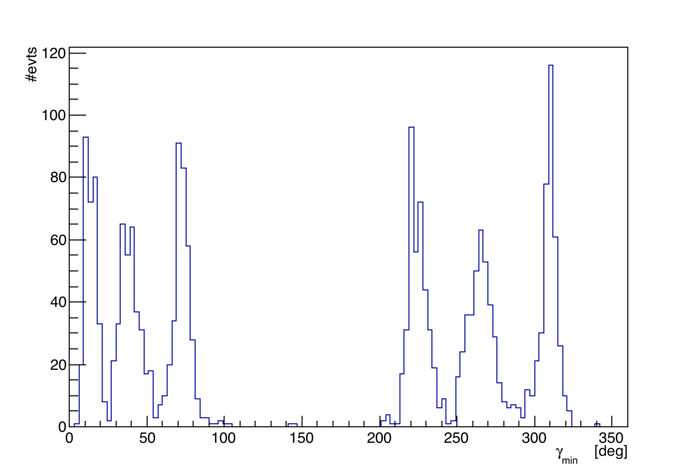}
\caption{The minima found with five decay modes, with \aSU free to vary in the fit. For each of the 401 sets of random combinations of three points in the Dalitz plot, a $\chi^2$ scan for $\gamma$ is performed and the minima $\gamma_{\mathrm{min}}$ are found. The histogram shows the accumulation of the minima across all 401 scans.}    
	\label{minima5modesSym1_fit}
\end{figure}

The results of the procedure with five modes are shown in
Table~\ref{Minima_sym1_5modes},
giving the central values ($\mu$),
asymmetric experimental uncertainties ($\sigma_L$, $\sigma_R$),
and the recomputed systematic uncertainty due to poorly resolved minima
($|\mu - \mu^{\rm{all}}|$).
The systematic uncertainty associated with \SU breaking is
also given in the table; this is the same as before by construction. The distribution of the minima across the scans is shown in Figure~\ref{minima5modesSym1_fit}.
The results for the minima are compatible
with the ones obtained with four modes.

\begin{table}[bthp]
\centering
\caption{The rates at which the different minima are obtained with five decay modes. A total of 401 scans are used.}
\label{freq_5modes}
\begin{tabular}{lcc}
 & Count  & Fraction (\%)  \\
\midrule
minimum 1	 & 	306	     & 	 76.3   \\
minimum 2	 & 	329	     & 	 82.0   \\
minimum 3	 & 	372	     & 	 92.3	\\
minimum 4	 & 	383	     & 	 95.5	 \\
minimum 5	 & 	378	     & 	 94.3	 \\
minimum 6	 & 	391	     & 	 97.5	 \\
\end{tabular}
\end{table}

\begin{table}[h!]
\caption{
  The minima found with five decay modes, allowing $\aSU$
  to vary in the fit.
  For each minimum, the central value for $\gamma$ is given ($\mu$),
  along with the asymmetric experimental uncertainty on the left-
  and right-hand sides ($\sigma_L$, $\sigma_R$).
  The quantities
 $|\mu-\mu^{\rm{all}}|$ and $|\mu^{4\,\rm{modes}}-\mu^{5\,\rm{modes}}|$
  are taken as estimates of the systematic uncertainties
  due to
  poorly resolved minima and flavour \SU breaking, respectively.
}
\label{Minima_sym1_5modes}
\begin{center}
\begin{tabular}{crrrcc}
&
\multicolumn{1}{c}{$\mu$} &
\multicolumn{1}{c}{$\sigma_L$} &
\multicolumn{1}{c}{$\sigma_R$} &
$|\mu-\mu^{all}|$ &
$|\mu^{4\,\rm{modes}}-\mu^{5\,\rm{modes}}|$ \\
\midrule
Minimum 1 & $ 11.9\degree$ & $ 5.8\degree$ & $ 9.1$ & $1.3$ & $1.0$ \\
Minimum 2 & $ 39.2\degree$ & $ 6.3\degree$ & $ 6.7$ & $1.2$ & $2.6$ \\
Minimum 3 & $ 71.3\degree$ & $ 9.5\degree$ & $ 9.3$ & $0.4$ & $2.4$ \\
Minimum 4 & $223.9\degree$ & $ 7.4\degree$ & $ 9.5$ & $0.1$ & $0.7$ \\
Minimum 5 & $265.0\degree$ & $11.0\degree$ & $10.0$ & $1.2$ & $1.3$ \\
Minimum 6 & $308.4\degree$ & $ 8.8\degree$ & $ 7.0$ & $0.6$ & $0.9$ \\
\end{tabular}
\end{center}
\end{table}

\end{appendices}

\clearpage
\bibliography{biblio}
\bibliographystyle{utphys}

\end{document}